\documentclass[11pt]{article}
\usepackage{showkeys}
\usepackage{amsmath,amssymb,cite}
\newcommand{\be}{\begin{equation}}
\newcommand{\ee}{\end{equation}}
\newcommand{\bea}{\begin{eqnarray}}
\newcommand{\eea}{\end{eqnarray}}
\newcommand{\pp}{~.}
\newcommand{\vv}{~,}
\newcommand{\cc}{{\mathbb{C}}}
\newcommand{\rr}{{\mathbb{R}}}

\numberwithin{equation}{section}
\usepackage{epsfig}

\begin{document}

\title{{\bf{\Large Twisted Galilean symmetry and \\ the Pauli 
principle at low energies}}}
\author{
{\bf {\normalsize Biswajit Chakraborty}$^{a,b,}
$\thanks{biswajit@bose.res.in}},\,
{\bf {\normalsize Sunandan Gan{g}opadhyay}$^{a,}
$\thanks{sunandan@bose.res.in}}\\
{\bf {\normalsize Arindam Ghosh Hazra}
$^{a,}$\thanks{arindamg@bose.res.in}},\, 
{\bf {\normalsize Frederik. G. Scholtz}
$^{b,a,}$\thanks{fgs@sun.ac.za}}\\
$^{a}$ {\normalsize S.~N.~Bose National Centre for Basic Sciences,}\\
{\normalsize JD Block, Sector III, Salt Lake, Kolkata-700098, India}\\[0.3cm]
$^{b}$ {\normalsize Institute of Theoretical Physics, University of
Stellenbosch,}\\{\normalsize Stellenbosch 7600, South Africa}\\[0.3cm]
}

\maketitle

\begin{abstract}

We show the twisted Galilean invariance of the noncommutative
parameter, even in presence of space-time noncommutativity.
We then obtain the deformed algebra of the Schr\"odinger field in 
configuration and momentum space by studying the action of the 
twisted Galilean group on the non-relativistic limit of the 
Klein-Gordon field. Using this deformed algebra we compute 
the two particle correlation function to study the possible 
extent to which the previously proposed violation of the Pauli 
principle may impact at low energies. It is concluded that any 
possible effect is probably well beyond detection at current energies.     

\vskip 0.2cm
{{\bf {Keywords}}: Noncomutative geometry, Twisted Galilean symmetry} 
\\[0.3cm]
{\bf PACS:} 11.10.Nx 

\end{abstract}

\section{Introduction}
The study of noncommutative (NC) geometry and its implications have gained
considerable importance in recent times as these studies are motivated
both from string theory \cite{sw} as well as from certain condensed
matter systems like the quantum Hall effect \cite{duval, dayi, fgs}.
 Here the canonical
NC structure is given by the following operator 
valued space-time coordinates,
\begin{equation}
\left[x^{\mu}_{op} , x^{\nu}_{op}\right] = i\theta^{\mu \nu}.
\label{1a} 
\end{equation}
Instead of working with functions of these operator-valued coordinates, one can
alternatively work with functions of $c$-numbered coordinates provided
one composes the functions through the $\star$ product defined as \cite{szabo}
\begin{eqnarray}
\alpha \ast_\theta \beta \;(x) &=& \left[ \alpha \exp\left(\frac{i}{2}
\overleftarrow{\partial_\mu} \theta^{\mu \nu}
\overrightarrow{\partial_\nu} \right) \beta \right] (x) 
\label{starprod} \\
\theta^{\mu \nu}&=&-\theta^{\nu \mu} \in \rr \vv \   x=
(x^0,x^1,\ldots,x^d) \pp \nonumber
\end{eqnarray}
The Poincar\'{e} group $\mathcal{P}$ or the 
diffeomorphism group $\mathcal{D}$ which 
acts on the NC
space-time $\rr^{d+1}$ defines a natural action on smooth functions 
$\alpha \in C^{\infty}(\rr^{d+1})$ as
\begin{equation}
\label{gaction}
( g \alpha) (x) = \alpha (g^{-1} x),
\end{equation}
for $g \in \mathcal{P}$ or $\in \mathcal{D}$.
However, in general 
\begin{equation}
(g \alpha) \ast_\theta (g \beta) \neq g (\alpha \ast_\theta \beta)
\label{gastarb} 
\end{equation}
showing that the action of the group $\mathcal{P}$ or $\mathcal{D}$ is not
an automorphism of the algebra ${\cal A}_\theta(\rr^{d+1})$, unless
one considers the translational sub-group.
This violation of Poincar\'e symmetry in particular is accompanied
by the violation of microcausality, 
spin statistics and CPT theorem
in general \cite{szabo, greenberg}. These results, which 
follow from the basic axioms
in the canonical (commutative) Quantum Field Theory (QFT), are no longer
satisfied in presence of noncommutativity in the manner discussed
above. Besides, NC field theories are afflicted with 
infra-red/ultra-violet (IR/UV) mixing. It is however possible for  some of 
these results to still go through
even after postulating weaker versions of the axioms used in 
the standard QFT. For example one can consider the proof
of CPT theorem given by Alvarez-Gaume et.al \cite{alvarez} where
they consider the breaking of Lorentz symmetry down to the subgroup
$O(1,1) \times SO(2)$, and replace the usual causal structure,
given by the light cone, by the light-wedge associated with the
$O(1,1)$ factor of kinematical symmetry group. One can also consider
the derivation of CPT and Spin-Statistics theorems by Franco et.al
\cite{franco} where they invoke only ``asymptotic commutativity''
i.e. assuming that the fields to be commuting at sufficiently large
spatial separations.

As all these problems basically stemmed from the above
mentioned non-invariance (\ref{gastarb}), it is desirable
to look for some way to restore the invariance. Indeed, 
as has been discovered by \cite{chaichian} 
and \cite{aschieri,Dimitrijevic:2004rf} (see
also the prior work of \cite{Oeckl:2000eg}) this invariance
is restored by introducing  a deformed coproduct, 
thereby modifying the corresponding Hopf algebra. 
Since then,  this deformed or twisted coproduct has been used 
extensively in the framework of relativistic quantum field theory,
as this approach seems to be  quite promising.

Two interesting consequences follow from the twisted 
implementation of the Poincar\'{e} group. 
 The first is that there is apparently no longer any IR/UV mixing 
\cite{bal1}, suggesting that the high and low energy 
sectors decouple, in contrast to 
the untwisted formulation.  The second striking consequence
 is an apparent violation of Pauli's principle \cite{bal}. 
This seems to be unavoidable if one wants to restore 
Poincar\'{e} invariance through the twisted coproduct. 
If there is no IR/UV mixing, 
one would expect that any violation of Pauli's 
principle would impact in either the high 
or low energy sector.  Experimental observation 
at present energies seems to rule out any 
effect at low energies, therefore if this picture is a 
true description of nature, we expect 
that any violation of the Pauli principle can only appear at 
high energies. It does, 
therefore, seem worthwhile as a consistency check 
to investigate this question in more 
detail and to establish precisely what the possible 
impact may be at low energies and why 
it may not be observable. One of the quantities where 
spin statistics manifests itself very 
explicitly is the two particle correlation function.  
A way of addressing this issue would 
therefore be to study the low temperature limit of the 
two particle correlation function in 
a twisted implementation of the Poincar\'{e} group. Since we 
are at low energies it would, however,
 be sufficient to study the non-relativistic limit, i.e., 
the Galilean symmetry. We therefore need
 to consider the question whether the Galilean symmetry can 
also be restored by a suitable twist 
of the coproduct. This is a non trivial point that 
requires careful investigation, 
as the Galilean algebra admits a central extension, in the form of mass,
unlike the Poincar\'{e} case and the boost generator does not
have a well defined coproduct action.
It may be recalled, in this context, 
that the presence of space-time
non-commutativity spoils the NC structure under Galileo
boost. This question is  all the more important because of  
the observation made by \cite{doplicher} that 
the presence of space-time non-commutativity  does not spoil the unitarity
of the NC theory.
However, we show that the presence of space-time non-commutativity
in the relativistic case does not have a well defined non-relativistic
($c \to \infty$) limit. Furthermore, space-time non-commutativity
gives rise to certain operator ordering ambiguities rendering
the extraction of a non-relativistic field in the 
$c \to \infty$ limit non trivial.

This paper is organised as follows. We discuss mathematical
preliminaries introducing the concept of Hopf algebra 
and the deformed or twisted coproduct in section 2.
Section 3 deals with a brief review of the twisted Lorentz
transformation properties of quantum space-time 
in sub-section (3.1), as was
discussed by \cite{chaichian, wess}. This is then extended
to the non-relativistic case in sub-section (3.2).
We then discuss briefly the  non-relativistic reduction
of the Klein-Gordon field to the Schr\"{o}dinger field in
(2 + 1) dimensions in commutative space in section 4,
which is then used to obtain the action of the twisted 
Galilean transformation on the Fourier coefficients in section 5.
We eventually obtain the action of twisted 
Galilean transformation on non-relativistic Schr\"{o}dinger fields
in section 6. In section 7 we discuss the implications of the subsequent deformed 
commutation relations on the two particle correlation function of a free gas in two spatial dimensions. We conclude in section 8. Finally, we have added
an Appendix where we have included some important aspects
of Wigner-In$\ddot{o}$nu group contraction in this context (i.e. 
Poincar\'{e} $\to$ Galileo), which we have made use of in the main text.

\section{Mathematical preliminaries}
In this section we give a brief review of the essential 
results in \cite{bal} for the purpose of application in later sections.

\noindent
Suppose that a group $G$ acts on a complex vector space $V$ by a
representation $\rho$. We denote this action by
\begin{equation}
v \rightarrow \rho(g) v 
\label{rhov}
\end{equation}
for $g \in G$ and $v \in V$. Then the group algebra $G^*$ also acts on
$V$.  A typical element of $G^*$ is
\begin{equation}
\int dg \,\alpha(g)\, g, \,\,\,\,\, \alpha(g) \in \cc 
\label{llll}
\end{equation}
where $dg$ is an invariant measure on $G$. Its action is
\begin{equation}
v \rightarrow \int dg \,\alpha(g) \, \rho (g) \, v \pp
\label{actgstar}
\end{equation}

Both $G$ and $G^*$ act on $V \otimes_\cc V$, the tensor product of
$V$'s over $\cc$, as well. These actions are usually taken to be
\begin{equation}
v_1 \otimes v_2 \rightarrow \left[ \rho(g) \otimes \rho(g) \right]
(v_1 \otimes v_2 ) = \rho(g) v_1 \otimes \rho(g) v_2 
\label{acttens}
\end{equation}
and
\begin{equation}
v_1 \otimes v_2 \rightarrow
\int dg \, \alpha(g) \, \rho(g) v_1 \otimes \rho(g) v_2 
\label{acens}
\end{equation}
respectively, for $v_1, v_2 \in V$.

In Hopf algebra theory \cite{majid, chai}, the action of 
$G$ and $G^*$ on tensor products
is defined by the coproduct $\Delta_{0}$, a homomorphism from $G^*$
to $G^* \otimes G^*$, which on restriction to $G$ gives a homomorphism
from $G$ to $G^* \otimes G^*$. This restriction specifies $\Delta_{0}$ on
all of $G^*$ by linearity. Hence, if
\begin{eqnarray}  
&& \Delta_{0}: \, g \rightarrow \Delta_{0}(g)  \\ \nonumber
&& \Delta_{0}(g_1)\Delta_{0}(g_2)=\Delta_{0}(g_1g_2)  
\label{xev}
\end{eqnarray}  
we have
\begin{equation}
\Delta_{0} \left(\int dg \, \alpha(g) \, g \right) = \int dg \, \alpha(g)
\, \Delta_{0}(g) \pp 
\label{pml}
\end{equation}
Suppose next
that $V$ is an algebra ${\cal A}$ (over $\cc$).
As ${\cal A}$ is an algebra, we have a rule for taking products of
elements of ${\cal A}$. That means that there is a multiplication map
\begin{eqnarray}
m: {\cal A} \otimes {\cal A} \rightarrow {\cal A}  \\
\alpha \otimes \beta
\rightarrow m (\alpha \otimes \beta)  \nonumber 
\label{multmap}
\end{eqnarray}
for $\alpha,\beta \in {\cal A}$, the product $\alpha \beta$ being $m
(\alpha \otimes \beta)$.

It is now essential that $\Delta_{0}$ be compatible with $m$, so that: 
\begin{equation}
m \left( (\rho \otimes \rho) \Delta_{0}(g) \left(\alpha \otimes \beta
\right)\right)=
\rho(g) m ( \alpha \otimes \beta) \pp
\label{compatib}
\end{equation}
In the Moyal plane, the multiplication denoted by the
map $m_{\theta}$, is NC and depends on
$\theta^{\mu \nu}$. It is defined by \footnote{The signature
we are using is $(+ , - , - , ...)$.}
\begin{eqnarray}
m_\theta ( \alpha \otimes \beta) &=& m_0 \left( e^{-\frac{i}{2} (i
\partial_\mu) \theta^{\mu \nu} \otimes (i \partial_\nu) } \alpha
\otimes \beta \right)\nonumber \\
&=& m_0 \left( F_\theta \alpha \otimes \beta
\right)\,
\label{multmoyal}
\end{eqnarray}
where $m_{0}$ is the point-wise multiplication of two functions and
$F_\theta$ is the twist element given by,
\begin{eqnarray}
F_\theta &=& e^{-\frac{i}{2}\theta^{\mu \nu} P_{\mu} \otimes P_{\nu}}
\nonumber \\
&=& e^{-\frac{i}{2} (i \partial_\mu) \theta^{\mu \nu}
\otimes (i \partial_\nu) } \quad ; P_{\mu} = i\partial_\mu.
\label{ftheta}
\end{eqnarray}
The twist element $F_\theta$ changes the coproduct to
\begin{eqnarray}
\Delta_{0} (g) \rightarrow \Delta_\theta (g) = 
 \hat{F}^{-1}_\theta \Delta_{0} (g) \hat{F}_\theta
\label{newcoprod}
\end{eqnarray}
in order to maintain compatibility with $m_{\theta}$, as
can be easily checked.
\noindent
In the case of the Poincar\'{e} group, if $\exp( i P {\cdot} a) $ is a
translation,
\begin{equation}
(\rho \otimes \rho)\Delta_\theta \left( e^{i P {\cdot} a} \right) 
e_p \otimes e_q = e^{i (p+q) {\cdot} a}  e_p \otimes
e_q\vv  \quad  \ \ (e_p(x) = e^{-ip\cdot x})  
\label{planetrans}
\end{equation}
while if $\Lambda$ is a Lorentz transformation
\begin{equation}
(\rho \otimes \rho)\Delta_\theta(\Lambda) e_p \otimes e_q
= \left[e^{\frac{i}{2}(\Lambda p)_\mu \theta^{\mu \nu}
    (\Lambda q)_\nu } e^{-\frac{i}{2} p_\mu \theta^{\mu \nu}
    q_\nu } \right] e_{\Lambda p} \otimes e_{\Lambda q} \pp 
\label{lorentz}
\end{equation}
These relations are derived in \cite{bal}.
Finally, let us mention the action of the coproduct $\Delta_{0}$ on 
the elements of a Lie-algebra $\cal{A}$ . 
The coproduct is defined on $\cal{A}$ by
\begin{equation}
\Delta_{0}(X) = X \otimes 1 + 1 \otimes X.
\label{cop}
\end{equation}
Its action on the elements of the corresponding 
universal covering algebra $\cal{U}(\cal{P})$
can be calculated through the homomorphism \cite{varily} i.e.
\begin{equation}
\Delta_{0}(XY) = \Delta_{0}(X) \Delta_{0}(Y) = XY \otimes 1 + X \otimes Y 
+ Y \otimes X +  1 \otimes XY.
\label{co}
\end{equation}
One can also easily check that this action of the coproduct on
the Lie-algebra is consistent with the action on the group element
defined by
\begin{equation}
\Delta_{0}(g) = g \otimes g.
\label{copg}
\end{equation}

\section{Transformation properties of tensors under space-time transformation}
\subsection{Lorentz transformation}
To set the scene for the rest of the paper, we give a brief review of the
Lorentz transformation properties in the commutative case in this subsection.
This, as we shall see,  turns out to be essential in understanding
the action of the Lorentz generators on any vector or tensor field.
Consider an infinitesimal Lorentz transformation
\begin{eqnarray}
x^{\mu}\rightarrow x^{\prime\mu}=x^{\mu}+\omega^{\mu\nu}x_{\nu}
\label{1}
\end{eqnarray}
where, $\omega^{\mu\nu}$ is an infinitesimal constant 
($\omega^{\mu\nu}=-\omega^{\nu\mu}$).
Under this transformation, any vector field $A_{\mu}$ transforms as
\begin{eqnarray}
A_{\mu}\rightarrow A_{\mu}^{\prime}(x^\prime)=A_{\mu}(x)
+{\omega_{\mu}}^{\lambda}A_{\lambda}(x)
\label{2}
\end{eqnarray}
Hence the functional change in $A_{\mu}(x)$ reads
\begin{eqnarray}
\delta_{0}A_{\mu}(x)&=&A_{\mu}^{\prime}(x)-A_{\mu}(x)\nonumber\\
&=&\omega^{\nu\lambda}x_{\nu}\partial_{\lambda}A_{\mu}(x)+\omega_{\mu\nu}A^{\nu}
\nonumber\\
&=& - \frac{i}{2}\omega^{\nu\lambda}J_{\nu\lambda}A_{\mu}
\label{3}
\end{eqnarray}
where, $J_{\nu\lambda}=  M_{\nu\lambda}+S_{\nu\lambda}$ are the total
Lorentz generators with $M_{\mu\nu}$ and  $S_{\mu\nu}$ 
identified with orbital and spin parts, respectively. This immediately
leads to the representation of $M_{\nu\lambda}$
\begin{eqnarray}
M_{\nu\lambda}= i(x_{\nu}\partial_{\lambda}-x_{\lambda}\partial_{\nu})
=  \left(x_{\nu}P_{\lambda}-x_{\lambda}P_{\nu}\right); 
\quad P_{\lambda} = i\partial_{\lambda}\,.
\label{4}
\end{eqnarray}
To find the representation of $S_{\nu\lambda}$, we make use of the relation
$\frac{i}{2}\omega^{\rho\lambda}(S_{\rho\lambda}A)_{\mu} = \omega_{\mu\nu}A^{\nu}$
obtained by comparing both sides of (\ref{3}). This leads to
\begin{eqnarray}
(S_{\alpha\beta})_{\mu\nu} = i(\eta_{\mu\alpha}\eta_{\nu\beta}-
\eta_{\mu\beta}\eta_{\nu\alpha})\,.
\label{5}
\end{eqnarray}
It can now be easily checked that $M_{\mu\nu},\ S_{\mu\nu}$ and
$J_{\mu\nu}$ all satisfy the same homogeneous Lorentz algebra
$SO(1 , 3)$:
\begin{eqnarray}
\left[M_{\mu\nu} , M_{\lambda\rho}\right] = i\left(
\eta_{\mu\lambda}M_{\nu\rho} - \eta_{\mu\rho}M_{\nu\lambda}
- \eta_{\nu\lambda}M_{\mu\rho} + \eta_{\nu\rho}M_{\mu\lambda}\right).
\label{5a}
\end{eqnarray}

\noindent
Setting $A_{\mu} = x_{\mu}$, where $x_{\mu}$ represents a
position coordinate of a spacetime point, yields
\begin{eqnarray}
\delta_{0}x_{\mu} = -\frac{i}{2}w^{\nu\lambda}(M_{\nu\lambda}
+S_{\nu\lambda})x_{\mu}=0
\label{6}
\end{eqnarray}
as expected, since the Lie derivative of the ``radial'' vector field
$\vec{X} = x^{\mu}\partial_{\mu}$ w.r.t. the ``rotation''
generators (\ref{4}) $M_{\mu\nu}$ vanishes i.e. 
${\cal{L}}_{M_{\mu\nu}}\vec{X} = 0 $.

\noindent 
Now we observe that the change in $x_{\mu}$ (not the functional change
$\delta_{0}x_{\mu}$ as in (\ref{3}))
defined by
\begin{eqnarray}
\delta x_{\mu} = x^{\prime}_{\mu}-x_{\mu} = {\omega_{\mu}}^{\nu}x_{\nu}
\label{6a}
\end{eqnarray}
can be identified as the action of $S_{\nu\lambda}$
on $x_{\mu}$
\begin{eqnarray}
\delta x_{\mu} = {\omega_{\mu}}^{\nu}x_{\nu}
= - \frac{i}{2}\omega^{\nu\lambda}\left(S_{\nu\lambda}x\right)_{\mu}
\label{7q}
\end{eqnarray}
with the representation of $S_{\nu\lambda}$ given in (\ref{5}).
Using (\ref{6}), one can also obtain the action
of $M_{\nu\lambda}$ on $x_{\mu}$ 
\footnote{Note that $\delta A_{\mu}=A_{\mu}^{\prime}(x^\prime)-A_{\mu}(x)
={\omega_{\mu}}^{\lambda}A_{\lambda}(x)$ is not the functional change
and $\delta x_{\mu}$ in (\ref{7}) is
obtained by setting $A_{\mu}=x_{\mu}$.}
\begin{eqnarray}
\delta x_{\mu} = -\frac{i}{2}\omega^{\nu\lambda}M_{\nu\lambda}x_{\mu}.
\label{7}
\end{eqnarray}
One can generalise this to higher second rank tensors 
$f_{\rho\sigma}(x)=x_{\rho}x_{\sigma}$ as
\begin{eqnarray}
\delta \left(x_{\lambda} x_{\sigma}\right) =
\left(- \frac{i}{2} w^{\mu\nu} M_{\mu\nu}\right) 
\left(x_{\lambda} x_{\sigma}\right)
\label{13}
\end{eqnarray}
since we can write
\begin{eqnarray}
M_{\mu\nu}f_{\rho\sigma} &=& i(x_{\mu}\partial_{\nu}-x_{\nu}\partial_{\mu})
f_{\rho\sigma}\nonumber\\
&=&i(f_{\mu\sigma}\eta_{\nu\rho}-f_{\nu\sigma}\eta_{\mu\rho}+
f_{\rho\nu}\eta_{\mu\sigma}-f_{\rho\mu}\eta_{\nu\sigma})
\label{1b}
\end{eqnarray}
where we have made use of (\ref{4}). This indeed shows the 
covariant nature of the transformation properties
of $f_{\rho\sigma}$.

\noindent
We now review the corresponding covariance property in the 
NC case under the twisted coproduct of Lorentz generators
\cite{chaichian}, \cite{wess}.
The issue of violation of Lorentz symmetry in noncommutative
quantum field theories has been known for a long time,
since the field theories defined on a noncommutative
space-time obeying the commutation
relation (\ref{1a}) between the coordinate operators,
where $\theta_{\mu\nu}$ is treated as a constant antisymmetric matrix, are obviously
not Lorentz invariant. However, it has been shown \cite{chaichian} that
there exists a new kind of  twisted Poincar\'{e}
symmetry under which quantum field theories defined on noncommutative
space-time are still Poincar\'{e}  invariant.

\noindent
To generalise to the NC case, first note that 
the star product between two vectors
$x_{\mu}$ and $x_{\nu}$  given as $x_{\mu} \star x_{\nu}$ is
not symmetric, unlike in the commutative case. One can,
however, write this as
\begin{eqnarray}
x_{\mu} \star x_{\nu} = x_{\{\mu} \star x_{\nu\}} + \frac{i}{2}
\theta_{\mu \nu}
\label{14}
\end{eqnarray}
where the curly brackets $\{ \}$ denotes symmetrization in the indices
$\mu$ and $\nu$. 
This can be easily generalised to higher ranks, showing that every 
tensorial object of the form $(x_{\mu} \star x_{\nu} \star .....\star
x_{\sigma})$ can be written as a sum of symmetric tensors of equal
or lower rank, so that the basis representation is symmetric.
Consequently $f_{\rho \sigma}$ should be replaced by 
the symmetrized expression
$f^{\theta}_{\rho \sigma} = x_{\{\rho} \star x_{\sigma \}} = 
\frac{1}{2} (x_{\rho} \star x_{\sigma } + x_{\sigma } \star x_{\rho})$,
and correspondingly the action of the Lorentz generator should be 
applied through the twisted coproduct (\ref{newcoprod}),
\begin{eqnarray}
M^{\theta}_{\mu \nu} f^{\theta}_{\rho \sigma} &=& M^{\theta}_{\mu \nu} 
m_{\theta} \left(x_{\rho} 
\otimes x_{\sigma} \right) = m_{\theta}\left(\Delta_{\theta}\left(
M_{\mu \nu}\right)\left(x_{\rho} 
\otimes x_{\sigma} \right)\right) \nonumber \\
&=& i(f^{\theta}_{\mu\sigma}\eta_{\nu\rho} - f^{\theta}_{\nu\sigma}\eta_{\mu\rho} +
f^{\theta}_{\rho\nu}\eta_{\mu\sigma} - f^{\theta}_{\rho\mu}\eta_{\nu\sigma}).
\label{17}
\end{eqnarray}
In the above equation, $M^{\theta}_{\mu \nu}$ denotes the usual 
Lorentz generator, but with the action of a twisted coproduct.
In \cite{chaichian}, it was shown that $M^{\theta}_{\mu \nu}
(\theta^{\rho \sigma}) = 0$ , and 
\begin{eqnarray}
 M^{\theta}_{\mu \nu}\left(S^{2}_{t}\right) = 0 \ ;
\quad (S^{2}_{t} = x_{\sigma} \star x_{\sigma})
\label{1744}
\end{eqnarray}
i.e., the antisymmetric tensor $\theta^{\rho \sigma}$ is twisted-Poincar\'{e}
invariant.
\subsection{Twisted Galilean Invariance}
In this sub-section we extend the above twisted Poincar\'{e}
result to the corresponding non-relativistic case. To demonstrate the need for this, 
consider the Galilean boost transformation,
\begin{eqnarray}
t \rightarrow t^{\prime} &=& t \nonumber \\
x^{i} \rightarrow x^{\prime i} &=&  x^{i} - v^{i}t
\label{18}
\end{eqnarray}
applied in the NC Galilean space time having the
following NC structure 
\begin{eqnarray}
\left[t , x^i\right] = i\theta^{0 i}
\ \ \ ; \ \ \  \left[ x^i , x^j\right] = i \theta^{ij}.
\label{18kj}
\end{eqnarray}
The corresponding expression in the boosted frame is
given by,
\begin{eqnarray}
\left[t^{\prime} , x^{\prime i}\right] &=& \left[t , x^{i}\right]
= i \theta^{0 i}\nonumber \\
\left[x^{\prime i} , x^{\prime j}\right] &=& i \theta^{i j}
+  i  \left( \theta^{0 i} v^j - \theta^{0 j}v^i\right).
\label{18a}
\end{eqnarray}
This shows that the NC structure in the primed frame
does not preserve its structure unless space-time non commutativity
disappears i.e. $\theta^{0 i} = 0$.
\noindent
In this section we demonstrate that even in the presence of space-time non commutativity the Galilean symmetry can be restored through an appropriate twist.
To do this we consider a tangent vector field $\vec{A}(x) = 
A^{\mu}(x)\partial_{\mu}$, in  Galilean space-time. 
Under Galilean transformations (\ref{18}),
\begin{eqnarray}
\label{19}
A^{i}(x) \rightarrow A^{\prime i}(x^{\prime}) &=&
\frac{\partial x^{\prime i}}{\partial x^{\mu}} A^{\mu}(x)
= A^{i}(x) - v^{i}A^{0}(x)  \\
A^{0}(x) \rightarrow A^{\prime 0}(x^{\prime}) &=& A^{0}(x). \nonumber
\end{eqnarray}
From (\ref{19}) it follows that
\begin{eqnarray}
\delta_{0} A^{\mu}(x) &=& A^{\prime \mu}(x) - A^{\mu}(x) \nonumber \\
&=& i v^{j} \left(- i t \partial_{j} A^{\mu}(x) + 
i \delta^{\mu}_{j} A^{0}(x)\right) \nonumber \\
&=& i v^{j} K_{j} A^{\mu}(x)
\label{20}
\end{eqnarray}
where
\begin{eqnarray}
 K_{j} A^{\mu}(x) &=& \left(- i t \partial_{j} A^{\mu}(x) + 
i \delta^{\mu}_{j} A^{0}(x)\right) \nonumber \\
&=& - tP_{j} A^{\mu}(x) +  i \delta^{\mu}_{j} A^{0}(x).
\label{21}
\end{eqnarray}
Setting $A^{\mu}(x) = x^{\mu}$\footnote{Here we identify $x^0$
to be just the time $t$, rather than $ct$.} we  easily see that 
$ K_{j} x^{\mu} = 0$, from which we get
\begin{eqnarray}
\delta x^{\mu} = i v^{j} t P_{j} x^{\mu} = i v^{j} K^{(0)}_{j} x^{\mu}
\label{22}
\end{eqnarray}
where, $K^{(0)}_{j} =  t P_{j}$.  This is the counterpart of (\ref{7}) 
in the Galilean case. 
In other words, here $K^{(0)}_{j}$ plays the same role as
$M_{\mu \nu}$ in the relativistic case.
Indeed, one can check that at the commutative level it has 
its own coproduct action
\begin{eqnarray}
K^{(0)}_{j} m\left( x^{\mu} \otimes x^{\nu} \right) = m \left(\Delta_{0}
\left(K^{(0)}_{j} \right)\left(x^{\mu} \otimes x^{\nu} \right)\right).
\label{23}
\end{eqnarray}
Here $K^{(0)}_{j}$ is clearly the boost generator 
$K^{(M)}_{j}$ (see equation \ref{12c} in Appendix) with $M = 0$. 
Note that with $M \neq 0$,
$K^{(M)}_{j}$ does not have the right coproduct action (\ref{23}). 
This is also quite satisfactory
from the point of view that the non-commutativity of space-time
is an intrinsic property and should have no bearing on the mass of
the system inhabiting  it. We also point out another dissimilarity
between the relativistic and non-relativistic case. In the  relativistic
case, the generators $M_{\mu\nu}$ (\ref{4}) can be regarded as
the vector field whose integral curve generates the Rindler
trajectories, i.e. the space-time trajectories of uniformly
accelerated particle. On the other hand, the vector field associated
with the parabolic trajectories of uniformly accelerated particle
in the non-relativistic case is given by $K^{NR}_{i}$ (\ref{12}),
which however cannot be identified with the Galileo boost
generator $K^{(M)}_{j}$  (\ref{12c}) (see Appendix), unlike
$M_{\mu\nu}$ in the relativistic case.

\noindent
At the NC level the 
action of the Galilean generator
should be applied through the twisted coproduct
\begin{eqnarray}
K^{\theta (0)}_{j} m_{\theta}\left( x^{\mu} \otimes x^{\nu} \right) =
 m_{\theta} \left(\Delta_{\theta}
\left(K^{(0)}_{j} \right)\left(x^{\mu} \otimes x^{\nu} \right)\right)\,.
\label{24}
\end{eqnarray}
Using this and noting  $\Delta_{\theta} \left(K^{(0)}_{j} \right)
=  \Delta_{0} \left(K^{(0)}_{j} \right)$ we have
\begin{eqnarray}
K^{\theta (0)}_{j} m_{\theta}\left( x^{\mu} \otimes x^{\nu} \right) &=& 
it\left(x^{\mu} \delta^{\nu}_{j} + \delta^{\mu}_{j} 
x^{{\nu}}\right) \nonumber \\
\Rightarrow K^{\theta (0)}_{j} m_{\theta}\left( x^{\mu} \otimes x^{\nu} 
- x^{\nu} \otimes x^{\mu} \right)  &=&0 \nonumber \\
\Rightarrow K^{\theta (0)}_{j} \left(\theta^{{\mu}{\nu}}\right) &=& 0
\label{25}
\end{eqnarray}
i.e.\, the antisymmetric tensor $\theta^{{\mu}{\nu}}$ is invariant
under twisted Galilean boost. Since the rest of the Galileo
generators have the same form as that of the Poincar\'{e} generators,
discussed in the previous section, this shows the complete twisted
Galilean invariance of $\theta^{\mu \nu}$.
\section{Non-Relativistic reduction in commutative space}
In this section we discuss the non-relativistic 
reduction ($c \to \infty$) of the Klein-Gordon field
to the Schr\"{o}dinger field in 2+1 dimension\footnote{The
procedure of non-relativistic reduction holds for any
space-time dimension.}, as this is used
in the subsequent sections to derive the deformed algebra
of the Schr\"{o}dinger field both in the momentum and in 
the configuration space. The deformed algebra in the momentum
space for the Klein-Gordon field has already been derived in \cite{bal}.
Therefore it is advantageous to consider the non-relativistic limit
of such a deformed algebra.

\noindent
To facilitate the process of constructing the $c \to \infty$ limit,
we reintroduce the  speed of light `$c$' in appropriate places
from dimensional consideration, but we still work in the unit
$\hbar = 1$.
We consider the complex Klein-Gordon field, 
satisfying the Klein-Gordon equation
\begin{eqnarray}
\label{eq:45} 
\left(\frac{1}{c^2}\partial_{t}^2 - \nabla^2 
+ m^2c^2\right)\phi(x)=0.
\end{eqnarray}
This follows from the extremum condition of the Klein-Gordon
action,
\begin{eqnarray}
S = \int dt d^2{\bf{x}}\left[\frac{1}{c^2}\dot{\phi}^{\star}
\dot{\phi} - \phi^{\prime \star}\phi^{\prime} - 
c^2 m^2\phi^{\star}\phi\right]\,.
\label{26}
\end{eqnarray}

The Schr\"{o}dinger field is identified from the  Klein-Gordon
field by isolating the exponential factor involving rest mass energy
and eventually taking the limit $c \to \infty$.  
To that end we set
\begin{equation}
\label{eq:46} 
\phi (\vec{x},t) = \frac{e^{-imc^2t}}{\sqrt{2m}}\psi
(\vec{x},t)
\end{equation}
which yields from (\ref{eq:45}) the equation
\begin{equation}
\label{eq:47} 
-\frac{1}{2m}\nabla^2\psi =i \frac{\partial
\psi}{\partial t}-\frac{1}{2mc^2}\frac{\partial ^2\psi}{\partial
t^2}.
\end{equation}
This reduces to the Schr\"odinger equation of a free positive
energy particle
in the limit $c \to \infty$.
In this limit the action (\ref{26}) also yields the 
corresponding non-relativistic action  as
\begin{eqnarray}
S_{NR} = \int dt d^2x \, \psi^{\star}\left(i \partial_{0} + 
\frac{1}{2m}\nabla^2\right)\psi.
\label{27w}
\end{eqnarray}
The complex scalar field $\phi({\bf{x}})$ can be Fourier expanded as
\begin{eqnarray}
\phi(\vec{x} , t) = \int d\mu(k)c\left[a(k)e_{k} +  
b^{\dagger}(k)e_{-k}\right]
\label{27}
\end{eqnarray}
where, $d\mu(k) = \frac{d^2\vec{k}}{2k_0 (2\pi)^2}$ is the
Lorentz invariant measure and $e_{k} = e^{- i k.x} =
e^{- i\left( Et - \vec{k}\cdot \vec{x}\right)}$.
The well known equal time commutation relations between
$\phi$ and $\Pi_{\phi}$ can now be used to find the commutation
relation between $a_{k}$ and $a^{\dagger}_{k}$,\footnote{Note 
$k^{\mu} = \left(\frac{E}{c} , \vec{k}\right)$}
\begin{eqnarray}
\left[a(k) , a^{\dagger}(k^{\prime})\right] = 
(2\pi)^2 \frac{2 k_{0}}{c}\ \delta^2\left(
\vec{k} - \vec{k^{\prime}}\right)
\label{281}
\end{eqnarray}
and likewise for $b(k)$. In order to get the Fourier 
expansion of the field in the non-relativistic
case, we substitute (\ref{eq:46})
  in (\ref{27}), which in the limit
$c \rightarrow \infty$ yields
\begin{eqnarray}
\psi(\vec{x} , t) = \int \frac{d^2{\vec{k}}}{(2\pi)^2}
\frac{\tilde{c}(k)}{\sqrt{2m}} \tilde{e}_{k} 
= \int \frac{d^2{\vec{k}}}{(2\pi)^2}
c(k)\tilde{e}_{k} 
\label{29}
\end{eqnarray}
where, $\tilde{e}_{k} = e^{- i \frac{ \vert \vec{k} \vert^2 t}{2m}} 
e^{ i\vec{k}\cdot \vec{x}}$, $\tilde{c}(k) = \lim_{c \to \infty} a(k)$ and 
$c(k) = \frac{1}{\sqrt{2m}}\tilde{c}(k)$ are the Schr$\ddot{o}$dinger
modes.
As in (\ref{eq:47}) only the positive energy part survives
in the $c \to \infty$ limit, so that this limit 
effectively projects  the positive frequency part.
The commutation relation (\ref{281}) reduces in the non 
relativistic limit ($c \rightarrow \infty$) to
\begin{eqnarray}
\left[\tilde{c}(k) , \tilde{c}^{\dagger}(k^{\prime})\right] &=& 
(2\pi)^2 2m\ \delta^2\left(
\vec{k} - \vec{k^{\prime}}\right) \nonumber \\
\left[c(k) , c^{\dagger}(k^{\prime})\right]&=& 
(2\pi)^2 \ \delta^2\left(
\vec{k} -  \vec{k^{\prime}}\right)\,. 
\label{28}
\end{eqnarray}

\noindent
From (\ref{29}) and (\ref{28}), we obtain 
\begin{eqnarray}
\left[\psi(\vec{x} , t) \ ,\ \psi^{\dagger}(\vec{y} , t)\right] &=& 
 \delta^2\left(
\vec{x} -  \vec{y}\right)\,.
\label{30}
\end{eqnarray}

\section{Action of twisted Galilean transformation on Fourier coefficients}
Let us consider the Fourier expansion of the relativistic 
scalar field $\phi(\vec{x} , t)$
\begin{eqnarray}
\phi(\vec{x} , t) &=& \int d\mu(k)c \tilde{\phi}(k)e_{k}
\label{31}
\end{eqnarray}
where we have deliberately suppressed the negative
frequency part as it does not survive in the non-relativistic
limit $c \to \infty$, as we have seen in the previous section.
Considering the action of the Poincar\'{e} group elements on
$\phi$, we get
\begin{eqnarray}
\rho(\Lambda_{c}) \phi &=& \int d\mu(k)c\, \tilde{\phi}(k)
e_{\Lambda_{c} k} = \int d\mu(k)c\, \tilde{\phi}(\Lambda^{-1}_{c} k)
e_{k}  \\
\rho \left( e^{i P {\cdot} a} \right)  \phi &=&  \int d\mu(k)c\, 
e^{ik\cdot a}\tilde{\phi}(k)e_{k}\pp
\label{32}
\end{eqnarray}
Thus the representation $\tilde{\rho}$ of the Poincar\'{e}
group on $\tilde{\phi}(k)$ is specified by
\begin{eqnarray}
\left(\tilde{\rho}(\Lambda_{c}) \tilde{\phi} \right) (k) &=&
\tilde{\phi}(\Lambda_{c}^{-1} k)  \nonumber \\
\left(\tilde{\rho}\left( e^{i P {\cdot} a} \right) \tilde{\phi}\right) (k)
&=& e^{i k {\cdot} a} \tilde{\phi}(k) \pp
\label{33}
\end{eqnarray}
Here homogeneous Lorentz transformations have been labeled by
$\Lambda_{c}$. The corresponding Galilean transformations
will be labeled by $\Lambda_{\infty}$ in the $c \to \infty$ limit.

\noindent
If $\chi$ is another scalar 
field, with Fourier expansion given by
\begin{eqnarray}
\chi(\vec{x} , t) = \int d\mu(q) c\, \tilde{\chi}(q)e_{q}\,
\label{34}
\end{eqnarray}
the tensor product of fields $\phi$ and $\chi$ is given by
\begin{equation}
\phi \otimes \chi = \int d\mu(k)d\mu(q)c^2\, \tilde{\phi}(k)
\tilde{\chi}(q)e_k \otimes e_q
\label{35} \pp
\end{equation}
Using (\ref{lorentz}) one obtains the action of the
twisted Lorentz transformation on the above tensor
product of the fields
\begin{eqnarray}
\Delta_{\theta}(\Lambda_{c})(\phi \otimes \chi) =
\int d\mu(k)d\mu(q)c^2\, \tilde{\phi}(\Lambda^{-1}_{c}k)
\tilde{\chi}(\Lambda_{c}^{-1}q) e^{\frac{i}{2}k_{\mu}
\theta^{\mu \nu}q_{\nu}}e^{-\frac{i}{2}(\Lambda^{-1}_{c}k)_{\alpha}
\theta^{\alpha \beta}(\Lambda^{-1}_{c}q)_{\beta}}\left(e_k \otimes
e_q\right)\,.
\label{34f}
\end{eqnarray}
Substituting (\ref{eq:46}) in the above equation, one can
write
the corresponding action of the twisted Lorentz transformations
on the tensor product of fields $\psi$ and $\xi$ (here $\xi$ is
the counterpart of $\psi$ for the field $\chi$ as in (\ref{eq:46})) as
\begin{eqnarray}
\Delta_{\theta}(\Lambda_{c})\left(\psi \otimes \xi\right) &=& \int d\mu(k)d\mu(q)2m c^2\, 
\tilde{\phi}(\Lambda^{-1}_{c}k)
\tilde{\chi}(\Lambda^{-1}_{c}q) e^{\frac{i}{2}k_{i}
\theta^{i j}q_j}e^{-\frac{i}{2}(\Lambda^{-1}_{c}k)_{l}
\theta^{l n}(\Lambda^{-1}_{c}q)_{n}}\nonumber\\
&& \hspace {4.5cm}\times e^{-2i O(\frac{1}{c^2}, ....)}
\left(\tilde{e}_k \otimes
\tilde{e}_q\right) .
\label{36} 
\end{eqnarray}
Note that we have set $\theta^{0i} = 0$ in the right hand side
of the above equation. The underlying reason is that 
the substitution (\ref{eq:46}) can be carried out only in the
absence of space-time non-commutativity ($\theta^{0i} = 0$)
as this removes any operator ordering ambiguities in (\ref{eq:46}).
This should not, however, be regarded as a serious restriction
as theories with space-time noncommutativity do not represent
a low energy limit of string theory \cite{greenberg, gomis}

\noindent
Hence in the  limit $c \rightarrow \infty$,
we can deduce the action of the twisted Galilean transformations
($\Lambda_{\infty}$) on tensor products of the non-relativistic fields:
\begin{equation}
\Delta_{\theta}(\Lambda_{\infty})\left( \psi \otimes \xi\right)
 =  \int \frac{d^2 \vec{k}d^2 \vec{q}}{(2\pi)^4}\tilde{\psi}
(\Lambda^{-1}_{\infty}k)\tilde{\xi}(\Lambda^{-1}_{\infty}q)
e^{\frac{i}{2}m v_1\theta (k_2 - q_2)}\left(\tilde{e}_k
\otimes \tilde{e}_{q}\right)\,.
\label{37w} 
\end{equation}
Here we have  considered a boost along the $x^1$ direction with
velocity $v_1$ and $\tilde{\psi}(k) = \lim_{c \to \infty}\tilde{\phi}(k)$,
$\tilde{\xi}(q) = \lim_{c \to \infty}\tilde{\chi}(q) $.

\noindent
From the above, one deduces the action of the twisted 
Galilean transformations ($\Lambda_{\infty}$) on the Fourier
coefficients of the non-relativistic fields
\begin{equation}
\Delta_{\theta}(\Lambda_{\infty})\left( \tilde{\psi} \otimes \tilde{\xi}\right)
\left(k , q\right) =  \tilde{\psi} \left(\Lambda^{-1}_{\infty}k\right)
\tilde{\xi} \left(\Lambda^{-1}_{\infty}q\right)
e^{\frac{i}{2}m v_1 \theta (k_2 - q_2)}
\label{37} \pp
\end{equation}
One can now easily generalise the above result for the case of any arbitary
direction of boost as
\begin{equation}
\Delta_{\theta}(\Lambda_{\infty})\left( \tilde{\psi} \otimes \tilde{\xi}\right)
\left(k , q\right) =  \tilde{\psi} \left(\Lambda^{-1}_{\infty}k\right)
\tilde{\xi} \left(\Lambda^{-1}_{\infty}q\right)
e^{\frac{i}{2}m \theta \vec{v} \times  (\vec{k} - \vec{q})}
\label{373} \pp
\end{equation}

\section{Quantum Fields}
In this section, we discuss the action of twisted Galilean 
transformation on non-relativistic Schr\"{o}dinger fields. 
A free relativistic complex quantum field $\phi$ of mass $m$ can be
expanded in the NC plane (suppressing the negative 
frequency part) as
\begin{eqnarray}
\phi(\vec{x} , t) = \int d\mu(k) c\, d(k)e_k.
\label{371}
\end{eqnarray}
This is just the counterpart of (\ref{27}) where $a(k)$
has been replaced by $d(k)$\footnote{Note that 
$a(k) = \lim_{\theta \to 0}d(k)$}. \\
The deformation algebra involving $d(k)$ has already been
derived in \cite{bal}. In this paper, we  derive the 
deformation algebra for the non-relativistic case.
The non-relativistic limit of the complex Klein-Gordon field
has already been discussed in the earlier section and the
expansion is the following:
\begin{eqnarray}
\psi(\vec{x} , t) = \int \frac{d^2{\vec{k}}}{(2\pi)^2}
\frac{\tilde{u}(k)}{\sqrt{2m}} \tilde{e}_k 
= \int \frac{d^2{\vec{k}}}{(2\pi)^2}
u(k)\tilde{e}_k \ \ ;\ \ u(k) = \frac{1}{\sqrt{2m}}\tilde{u}(k)
\label{38}
\end{eqnarray}
where $\tilde{u}(k) = \lim_{c \to \infty} d(k)$. \\
\noindent
Note that 
$\tilde{c}(k), c(k)$ are the limits of the operators
$\tilde{u}(k) , u(k)$ respectively in the limit
$\theta^{\mu \nu} = 0$, and they satisfy the relations (\ref{28}).
We now argue that such relations are incompatible for $\theta^{\mu
\nu} \neq 0$. Rather, $u(k)$ and $u^\dagger(k)$
 fulfill certain deformed relations which reduce to (\ref{28})
 for $\theta^{\mu \nu} = 0$. \\
\noindent
Suppose  that
\begin{equation}
u(k) u(q) = \tilde{T}_\theta(k,q)
u(q) u(k)\,  
\label{40}
\end{equation}
where, $\tilde{T}_\theta$ is a $\cc$-valued function of $k$ and $q$
yet to be determined. The transformations of $u_{k} u_{l} = 
(u \otimes u) (k,l)$ and $u_{l} u_{k}$
 are determined by $\Delta_\theta$. 
Applying $\Delta_\theta$ on (\ref{40}) and using (\ref{37}),
we get the following\footnote{Without loss
of generality, we consider the boost to be along the $x^1$ direction
for calculational convenience. Also we set $v_1 = v$.}:
\begin{equation}
u \left(\Lambda^{-1}_{\infty}k\right)
u \left(\Lambda^{-1}_{\infty}q\right)
e^{\frac{i}{2}m v \theta (k_2 - q_2)}
= \tilde{T}_\theta(k,q) u \left(\Lambda^{-1}_{\infty}q\right)
u \left(\Lambda^{-1}_{\infty}k\right)
e^{\frac{i}{2}m v \theta (q_2 - k_2)}.
\label{41}
\end{equation}
Using (\ref{40}) again in the left hand side of (\ref{41}), 
we get:
\begin{equation}
\tilde{T}_\theta \left(\Lambda^{-1}_{\infty}k , \Lambda^{-1}_{\infty}q\right)
= \tilde{T}_\theta(k,q) e^{-i m v \theta (k_2 - q_2)}.
\label{42}
\end{equation}
Note that this equation can also be obtained from the corresponding
relativistic result \cite{bal} in the $c \to \infty$ limit provided one
takes $\theta^{0 i} = 0$ right from the beginning, otherwise 
the exponential factor become rapidly
oscillating in the  $c \to \infty$ limit, yielding no
well defined non-relativistic limit. Thus in the absence of 
space-time noncommutativity one has an appropriate non-relativistic
limit and  the above mentioned operator ordering ambiguities
can be avoided.\\
\noindent
The solution of (\ref{42}) is\footnote{Note that the non-relativistic
form of the twist element also appears in \cite{luk}.}
\begin{eqnarray}
\tilde{T}_\theta(k,q) = \eta e^{ i k_{i} \theta^{i j}  q_{j} }\ ; 
\quad (i, j = 1, 2)\,
\label{43}
\end{eqnarray}
where $\eta$ is a Galilean-invariant function and
approaches the value $\pm 1$ for bosonic and fermionic fields
respectively in the limit $\theta = 0$\footnote{The value
of $\eta$ can be actually taken to be $\pm 1$ for 
bosonic and fermionic fields for all $\theta^{\mu \nu}$ \cite{bal}. An exactly
similar non-relativistic reduction of the Dirac equation can also be done
for the fermionic case.}.
Therefore, substituting  (\ref{43}) in (\ref{40}) we finally have
\begin{equation}
u(k) u(q) = \eta e^{ i k_{i} \theta^{i j}  q_{j} }
u(q) u(k).
\label{44}
\end{equation}
The adjoint of (\ref{44}) gives:
\begin{equation}
u^\dagger(k) u^\dagger(q) = \eta e^{ i k_{i} \theta^{i j}  q_{j} }
u^\dagger(q) u^\dagger(k) .
\label{45}
\end{equation}
Finally the creation operator $u^\dagger(q)$ carries momentum $-q$, hence its
deformed relation reads:
\begin{equation}
u(k) u^{\dagger}(q) =  \eta e^{- i k_{i} \theta^{i j}  q_{j} } 
u^{\dagger}(q) u(k) + (2\pi)^2    \delta^2(k-q) .
\label{46}
\end{equation}
Using (\ref{44}) and (\ref{46}), one can easily obtain 
the deformation algebra involving  the non-relativistic fields
$\psi(x)$ in the configuration space:
\begin{eqnarray}
\psi(x)\psi(y) &=& \int d^2x^{\prime}d^2y^{\prime}
\Gamma_{\theta}(x, y, x^{\prime}, y^{\prime})
\psi(y^{\prime})\psi(x^{\prime}) \quad ;\ \ \theta \neq 0 \nonumber \\
\psi(x)\psi(y)&=& \eta \psi(y)\psi(x) \quad ;\ \ \theta = 0\, 
\label{46k}
\end{eqnarray}
\begin{eqnarray}
\psi(x)\psi^{\dagger}(y) &=& \int d^2x^{\prime}d^2y^{\prime}
\Gamma_{\theta}(x, y, x^{\prime}, y^{\prime})\psi^{\dagger}(y^{\prime})
\psi(x^{\prime}) + \delta^2(\vec{x} - \vec{y})\ \ 
;\ \theta \neq 0 \nonumber \\
\psi(x)\psi^{\dagger}(y) &=& \eta \psi^{\dagger}(y)\psi(x) + 
\delta^2(\vec{x} - \vec{y})
\quad ;\ \ \theta = 0\,
\label{46v}
\end{eqnarray}
where,
\begin{eqnarray}
\Gamma_{\theta}(x, y, x^{\prime}, y^{\prime}) 
= \frac{\eta}{(2\pi)^2}exp\left(\frac{i}{\theta}
\left[(x^{\prime}_1 - x_1)(y_2 - y^{\prime}_2) - 
(x^{\prime}_2 - x_2)(y_1 - y^{\prime}_1)\right]\right).
\end{eqnarray}
\section{Two particle correlation function}
In this section we compute the two particle correlation 
function for a free gas in 2+1 dimensions using the 
canonical ensemble, i.e., we are interested in 
the matrix elements $\frac{1}{Z}\langle r_1,r_2|e^{-\beta H}|
r_1,r_2\rangle$, where $Z$ is the canonical partition function
and $H$ is the non-relativistic Hamiltonian.  
The physical meaning of this function is quite simple; 
it tells us what the probability is to find particle two 
at position $r_2$, given that particle one is at $r_1$, i.e., 
it measures two particle correlations.  The relevant two 
particle state is given by
\begin{eqnarray}
|r_{1}, r_{2}\rangle&=&
\hat\psi^{\dag}(r_{1})\hat\psi^{\dag}(r_{2})|0\rangle\nonumber\\
&=&\int \frac{dq_{1}}{(2\pi)^2}\frac{dq_{2}}{(2\pi)^2}
e^{*}_{q_{1}}(r_{1})e^{*}_{q_{2}}(r_{2})u^{\dag}(q_{1})u^{\dag}(q_{2})
|0\rangle\,.
\label{rh1}
\end{eqnarray}
The two particle correlation function can therefore be written as
\begin{eqnarray}
\langle r_{1}, r_{2}|e^{-\beta H}|r_{1}, r_{2}\rangle
&=&\int dk_{1}dk_{2}e^{-\frac{\beta}{2m}(k^{2}_{1}+k^{2}_{2})}
|\langle r_{1}, r_{2}|k_{1}, k_{2}\rangle|^{2}
\label{rh2}
\end{eqnarray}
where we have introduced a complete set of momentum 
eigenstates $|k_{1}, k_{2}\rangle$.

Using (\ref{46}) and noting that
\begin{eqnarray}
|k_{1}, k_{2}\rangle=u^{\dag}(k_1)u^{\dag}(k_2)|0\rangle
\label{rh3}
\end{eqnarray}
we finally obtain
\begin{eqnarray}
C(r)\equiv \frac{1}{Z}\langle r_{1}, r_{2}|e^{-\beta H}|r_{1}, r_{2}\rangle
&=&\frac{1}{A^2}\left(1\pm\frac{1}{1+\frac{\theta^2}{\lambda^4}}
e^{- 2 \pi\, r^2/(\lambda^{2}(1+\frac{\theta^2}{\lambda^4}))}\right)
\label{rh4}
\end{eqnarray}
where $A$ is the area of the system and $\lambda$ is the mean 
thermal wavelength given by
\begin{eqnarray}
\lambda&=&\left(\frac{2\pi\beta}{m}\right)^{1/2}\quad;\quad
\beta=\frac{1}{k_{B}T}
\label{rh5}
\end{eqnarray}
and $r=r_{1}-r_{2}$. The plus and the minus signs indicate
bosons or fermions.

Although this calculation was done in 2+1 dimensions, it is 
clear that the result generalizes to higher dimensions by 
replacing $\theta^2$ by an appropriate sum of $(\theta^{ij})^2$. 
 The conclusions made below, based on the general structure 
of the correlation function, will therefore also 
hold in higher dimensions. 

As expected this result reduces to the standard (untwisted)
 result in the limit 
$\theta\rightarrow0$ \cite{pathria}.  
Furthermore it is immediately clear that when 
$\lambda>>\sqrt{\theta}$, i.e., in the low temperature limit, 
there is virtually no deviation from the untwisted result
 as summarized in figure \ref{corr}.  This is reassuring 
as it indicates that the implied violation of Pauli's 
principle will have no observable effect at current energies.
  Indeed, keeping in mind that $\sqrt{\theta}$ is probably at
 the Planck length scale any deviation will only become apparent
 at very high temperatures, where the non-relativistic limit 
is invalidated.  Note, however, that in contrast to the 
untwisted case the correlation function for fermions does not 
vanish in the limit $r\rightarrow 0$.  Thus, there is a 
finite probability that fermions may come very close to each other.
  Once again this probability is determined by $\theta$ 
and thus very small, which probably renders it undetectable.  
Due to this property of the twisted correlation function one 
also expects that the equation of state of a free fermion gas 
will be much softer at high densities than the untwisted one.  
This is most clearly seen from the exchange potential 
$V(r)=-k_BT\log C(r)$ \cite{pathria} shown 
in figure \ref{exc}. This clearly demonstrates the change 
from a hardcore potential in the untwisted case to a soft 
core potential in the twisted case.  This may have possible 
astrophysical implications, although it is dubious that these 
densities are even reachable in this case.  In any case the 
assumptions we made here are certainly violated at these extreme
 conditions and a much more careful analysis is required to 
investigate the high temperature and high density consequences
 of the twisted statistics. Another interesting point to note 
from figure \ref{exc} is that the twisted statistics has, 
even at these unrealistic values of $\frac{\theta}{\lambda^2}$,
 virtually no effect on the bosonic correlation function at 
short separation.  This probably suggests that there will be no 
observable effect in Bose-Einstein condensation experiments.  
These results may also have 
interesting consequences for condensed matter systems such 
as the quantum Hall effect where the NC parameter 
is related to the inverse magnetic field.

\setlength{\unitlength}{1mm}
\begin{figure}
\begin{picture}(53,53)
\put(30, 0){\epsfig{file=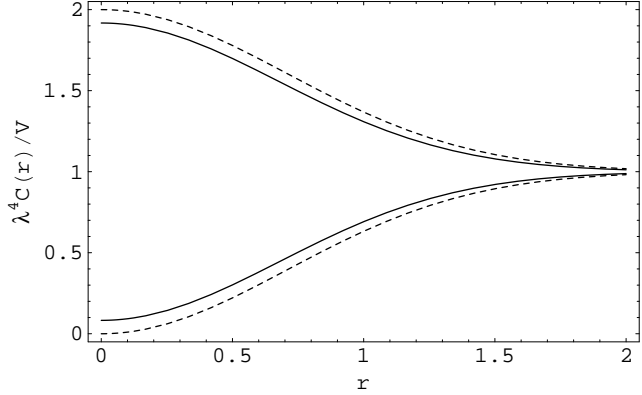, height=53mm}}
\end{picture}
\caption {Two particle correlation function $C(r)$.  
The upper two curves is the bosonic case and the 
lower curves the fermionic case.  The solid line shows the 
twisted result and the dashed line the untwisted case. This 
is shown for a schematic value of $\frac{\theta}{\lambda^2}=0.3$.  
The separation $r$ is measured in units of the thermal length $\lambda$.}
\label{corr}
\end{figure}

\setlength{\unitlength}{1mm}
\begin{figure}
\begin{picture}(53,53)
\put(30, 0){\epsfig{file=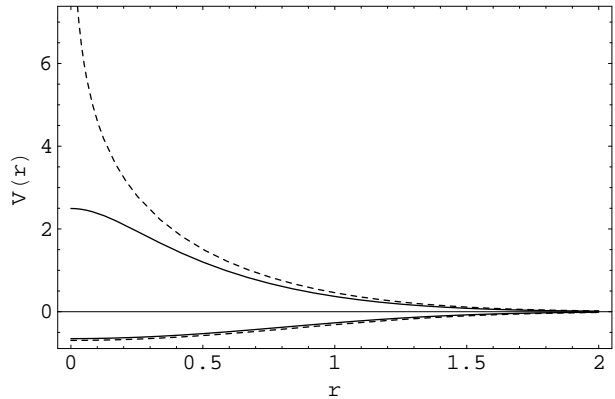, height=53mm}}
\end{picture}
\caption {Exchange potential $V(r)$ measured in units of $k_BT$.  The irrelevant additive constant has been set zero.  The upper two curves is 
the fermionic case and the lower curves the bosonic case.  The 
solid line shows the twisted result and the dashed line the untwisted 
case. This is shown for a schematic value of $\frac{\theta}{\lambda^2}=0.3$.  
The separation $r$ is measured in units
 of the thermal length $\lambda$.}
\label{exc}
\end{figure}
\section{Conclusions}
\label{concl}
We have shown that the NC parameter is twisted
Galilean invariant even in presence of space-time noncommutativity.
This is significant in view of the fact that the usual 
Galilean symmetry is spoiled in presence of space-time noncommutativity.

We have derived the deformed algebra of the Schr\"odinger field 
in configuration and momentum space.  
This was done by studying the action of the twisted 
Galilean symmetry on the Schr\"odinger field as 
obtained from a non-relativistic reduction of the Klein-Gordon field.
Here we had to consider the absence of any space-time noncommutativity
as, otherwise one can not define a proper non-relativistic limit.

The possible consequences of 
this deformation in terms of a violation of 
the Pauli principle was studied by computing the two 
particle correlation function. The conclusion is that 
any possible effect of the twisted statistics
 only show up at very high energies, while the effect 
at low energies should be very small, 
consistent with current experimental observations.  
Whether this effect will eventually be 
detectable through some very sensitive experiment is an open, 
but enormously interesting and 
challenging question.  
\section*{Acknowledgements}

This work was supported by a grant under the Indo--South African research
 agreement between the Department of Science and Technology,
Government of India and the South African National Research Foundation.
  FGS would like to thank the S.N. Bose National Center for
 Basic Sciences for their hospitality in the period that this
work was completed. BC would like to thank the Institute of Theoretical
Physics, Stellenbosch University for their hospitality during 
the period when part of this work was initiated. 
SG and AGH would like to thank Dr. Sachin Vaidya for some
useful discussion.

\appendix
\section{A brief derivation of Wigner-In$\ddot{o}$nu group contraction
of Poincar\'{e} group to Galilean group}
\label{appendix}
Here we summarise the well known Wigner-In$\ddot{o}$nu 
group contraction from Poincar\'{e} to Galilean algebra in order
to highlight some of the subtleties involved, as these have
direct bearings on the issues discussed in section (3.1).

\noindent
To begin with let us consider a particle undergoing uniform acceleration
`$a$', along the $x$ direction, measured in the instantaneous
 rest frame of the particle. A 
typical space-time Rindler trajectory is given by the hyperbola
\begin{eqnarray}
x^2 - c^2t^2 = {\rho}^2 
\label{8}
\end{eqnarray}
so that the acceleration $A(t)$ w.r.t the fixed observer with
the above associated coordinates ($t , x$) measured at time
$t$ is,
\begin{eqnarray}
A(t) = \frac{dV(t)}{dt} &=& \frac{c^2}{x}\left( \frac{\rho^2}{x^2}
 \right).\nonumber 
\end{eqnarray}
Since the frame $(x , t)$ appearing in (\ref{8}) coincides with that of the
fixed observer at time $t=0$, we must have
\begin{eqnarray}
\Rightarrow a = A(t = 0) &=& \frac{c^2}{\rho}  
\label{9}
\end{eqnarray}
where $\rho$ is
the distance measured at that instant from the origin.
To take the non-relativistic limit, we have to take both 
$c \rightarrow \infty$
and $\rho \rightarrow \infty$ such that $\frac{c^2}{\rho} = a$
is held constant.
For example, the corresponding non-relativistic expression 
$\bar{x}$ for the distance travelled by the particle in time $t$
is obtained by identifying
\begin{eqnarray}
\bar{x} = \lim_{{c \to \infty}  {\rho \to \infty} } 
\left( x - \rho \right) = \frac{1}{2} a t^2
\label{10}
\end{eqnarray}
which reproduces the standard result.

Now let us consider the Lorentz generator along the $x$ direction
$M_{01} = i \left(x_0 \partial_{1} - x_1 \partial_{0}\right)$.
This can be rewritten in terms of $\bar{x}$ using (\ref{10}),
\begin{eqnarray}
M_{01} &=& i c\left( t \frac{\partial}{\partial \bar{x}} + \frac{1}{a}
\left( 1 + \frac{\bar{x}}{\rho} \right)\frac{\partial}{\partial t}\right)
\nonumber \\
&=& cK_1 .
\label{11}
\end{eqnarray}
Note that $K_1$ by itself  does not have any $c$ dependence,
the non-relativistic limit of $K_1$ can thus be obtained by
just taking the limit $\rho \to \infty$, which yields
\begin{eqnarray}
K^{NR}_1 = \lim_{\rho \to \infty} K_1 =  
t \frac{\partial}{\partial \bar{x}} + \frac{1}{a}\frac{\partial}{\partial t}.
\label{12}
\end{eqnarray}
Although this vector field indeed generates the integral curve in
the $t$, $\bar{x}$ plane which is a parabola given by (\ref{10}),
it can not be identified with the Galileo boost generator because
\begin{eqnarray}
\left[ K^{NR}_i , K^{NR}_j \right] \sim \left( P_i - P_j\right).
\label{12b}
\end{eqnarray}
The Galilean algebra on the other hand is obtained by taking
the limit $c \to \infty$ of the commutators 
involving boost in the following way:
\begin{eqnarray}
\left[ \bar{K_{1}} , \bar{K_{2}} \right] &=& \lim_{c \to \infty} \frac{1}{c^2}
\left[ M_{01} , M_{02} \right] = \lim_{c \to \infty} \frac{1}{c^2} M_{12}
= 0 \nonumber \\
\left[ P_1 , \bar{K_{1}} \right] &=& \lim_{c \to \infty} \frac{1}{c}
\left[ P_1 , M_{01} \right] = \lim_{c \to \infty} \frac{1}{c^2} P_0
= i M \nonumber \\
\left[ \bar{K_{1}} , J \right] &=& \lim_{c \to \infty} \frac{1}{c}
\left[ M_{01} , M_{12} \right] = i \bar{K_{2}},
\label{12a}
\end{eqnarray}
\noindent
where $M$ is identified as the mass. The rest of the commutators
have the same form as that of Poincar\'{e} algebra.
This is nothing but the famous Wigner-In$\ddot{o}$nu group 
contraction, demonstrated here in construction of the
Galilean algebra as a suitable limit of the Poincar\'{e} algebra.\\
A simple inspection, at this stage, shows the following form of the
Galileo boost generators 
\begin{eqnarray}
\bar{K_{i}} = K^{(M)}_{i} = 
it\frac{\partial}{\partial \bar{x}_{i}} + M\bar{x}_{i}
\label{12c}
\end{eqnarray}
Clearly the rest of the generators in Galilean algebra 
have the same form as Poincar\'{e} algebra. For completeness 
we enlist the full Galilean algebra in (2 +1) dimension:
\begin{eqnarray}
\left[ K^{(M)}_{i} , K^{(M)}_{j} \right] &=&
\left[ P_i , P_j \right] =  \left[ P_i , H \right] = 
\left[ J , H \right] = 0 \nonumber \\
\left[ P_i , K^{(M)}_{j} \right] &=& i \delta_{ij} M\nonumber \\
\left[ P_i , J \right] &=& i \epsilon_{ij} P_{j}\nonumber \\
\left[ K^{(M)}_{i} , J \right] &=& i \epsilon_{ij} K^{(M)}_{j}\nonumber \\
\left[ P_i , M \right] &=&  \left[ H , M \right] = 
\left[ J , M \right] = \left[ K^{(M)}_{i} , M \right] = 0.
\label{12co}
\end{eqnarray}
Finally note that, here the mass $M$ plays the role of
 central extension of the centrally extended Galilean algebra.



\end{document}